# Room Temperature deposition of ZnO and Al:ZnO ultrathin films on glass and PET substrates by DC sputtering Technique.

[a]Mohammed RASHEED, [b]RÉGIS BARILLÉ

[a,b]MOLTECH-Anjou, Université d'Angers/UMR CNRS 6200, 2 Bd Lavoisier, 49045 Angers, France

corresponding email: rasheed.mohammed40@yahoo.com

## Abstract

In the present work, ultrathin films of Zinc oxide (ZnO) and Aluminum doped zinc oxide (AZO) with 20 nm thick were fabricated at 300 K on glass and Polyethylene terephthalate (PET) substrates by means of direct-current sputtering method. The structural morphology of the films were analyzed by (XRD) diffractometry. The average transmittance of films deposited on different substrates showed high transparency (over 80%) in the visible spectrum. The objective of the present work is to investigate the thicknesses and optical constants of ZnO and AZO ultrathin films prepared by DC sputtering onto (glass and PET) substrates using two methods (UV-vis-NIR spectrophotometer and ellipsometer (SE) by new amorphous dispersion formula) with comparison to check and confirm the results that obtained from UV with those obtained from SE measurements. The optical constants of films were extracted and compared using UV and SE techniques in the range of (200- 2200 nm) with increment of 1 nm in order to confirm the accuracy of the UV and shown an excellent agreement.

## 1. Introduction

Zinc oxide is a famous electro-optic material with a wide direct optical band gap semiconductor (of about 3.3-3.44 eV) at room temperature of the II-VI semiconductor group. This inorganic compound has a hexagonal wurtzite structure and large exciton binding energy of about 60 meV at room temperature [1-2]. ZnO have a high transmittance spectrum in the visible and near-ultraviolet range. These features makes it a good candidate for a wide range of applications such as solid state display devices, optical coatings heaters chemical sensors [3-4], hybrid and hetrojunction solar cells [5], light emitting diodes [6], bioimaging [7], and conductive transparent layer for photovoltaic applications [8-9]. Among the processes of fabrication for transparent zinc oxide thin films RF magnetron sputtering [10], spray pyrolysis [11], chemical vapor deposition (CVD) [4] pulsed laser deposition (PLD) [12], sol-gel technique [13], and DC sputtering technique [14] have been used. ZnO thin film component in transparent conducting oxides have been doped with various group II and group III metals ions (metal-doped oxides) such as aluminum (Al), indium (In), gallium (Ga), copper (Cu), cadmium, (Cd), manganese (Mn), magnesium (Mg), Iron (Fe) to enhance their surface morphology structures, optical and electrical properties. Doping is done to get stability, high conductivity, and high transparency. Aluminum doped zinc oxide is very sufficient for this purpose [1] is used as TCO layer (as anode) in organic and polymer LED structure. Most of these thin films are fabricated with amorphous or polycrystalline microstructures. TCOs used as thin-film electrodes in solar cells should have a minimum carrier concentration approximately 1020 cm$^{-3}$ for low resistivity and an optical energy gap greater than 3.3 eV at room temperature [1]. Many researchers have been interesting in using DC sputtering technique because it is not a line of sight method and can use diffusive spreading for coating and coat around the corners, high temperatures are not necessary and prevent PET flexible polymer plastic substrates from damage. High quality thin films on PET substrate give the possibility to be alternative substrates to the standard glasses. Al-doped ZnO films have good optical properties and low resistivity; exhibit a sharp UV cut-off, a high refractive index in the IR region and high transparency in the visible spectra. In addition, the optical band gap of zinc oxide can be enhanced by doping with aluminum [15].

At the present, we did not find any articles characterizing zinc oxide and Al doped Zinc oxide thin films fabricated by a DC sputtering technique and deposited on glass and PET substrates determining the optical constants of these thin films using (UV and SE) methods with the comparison between them.

ZnO and AZO ultrathin films were prepared by DC sputtering technique onto (glass and PET) substrates. The main goal of the present work is to investigate the optical constants of the films using two methods (UV and SE) with comparison between them.

## 2. Experimental

We made ZnO and ZnO:Al films by DC magnetron sputtering technique using Zn:Al (98%, 2%), the glass (~1.1 mm) and PET substrates (HIFI PMX739, ~175 µm) were placed on a rotating disk in a vertical direction with respect to the target and kept at ambient temperature. The distance between the target and the substrate was kept at about 7 cm, the deposition rate and the vacuum pressure of these films was: 4 nm/min and 5 Pa and (the deposition current I = 100 mA) with the sputtering power of 100W at room temperature for both. Commercial microscope slide plane glasses (25x20 mm) plates and PET used as substrates in this work. Glass substrates were cleaned carefully in an ultrasonic bath treatment with ethanol, acetone, deionized water and dichloromethane during 20 min and dried with nitrogen gas jet. Film thicknesses measured by a surface Profilometer with a Veeco 6M Metrology and SE measurements, the results are shown in table 1. It can be seen that the average error percentage between the exact and measured values of the film thicknesses was 3.9 %. The crystal structures of the samples deposited on glass and PET substrates were analyzed by D8 Advance Brucker diffractometer using a standard CuKα 1,2 (λ=0.15406 nm) and appeared amorphous.

### 2.4 Optical property measurements

The optical properties of films were obtained at room temperature in the wavelength by a PerkinElmer Lambda 950 (UV/Vis/NIR) Spectrophotometer and using a phase modulated spectroscopic ellipsometry type (UV-VISE-NIR Horiba Jobin Yvon) at $70^0$ angle of incident with increments of 1 nm at spectral range of (200-2200) nm. Practically, the ellipsometry described by the complex reflectance ratio ($\rho$) which is expressed by the ratio between the perpendicular (*s*-) and parallel (*p*-) polarized light reflection coefficients called (Fresnel reflection coefficients) expressing the electrical field alteration after reflection [16-17]:

$$\rho = \left|\frac{r_p}{r_s}\right| = tan\psi e^{i\Delta} \tag{1}$$

where $\psi$ is the amplitude ratio, $\Delta = \delta_p - \delta_s$ is the phase shift difference

New Amorphous dispersion formula can be expressed in the following equations [18-21]:

$$k(\omega) = \begin{cases} \frac{f_j.(\omega-\omega_g)^2}{(\omega-\omega_j)^2+\Gamma_j^2}, \ for \ \omega > \omega_g \\ 0, \ for \ \omega \leq \omega_g \end{cases} \quad and \quad n(\omega) = \begin{cases} n_\infty + \frac{B_j.(\omega-\omega_j)^2+c}{(\omega-\omega_j)^2-\Gamma_j^2}, \ for \ \omega > \omega_g \\ 0, \ for \ \omega \leq \omega_g \end{cases}$$

where $k(\omega)$ is the extinction coefficient, $n(\omega)$ is the refractive index, $B_j = \frac{f_j}{\Gamma_j}\left(\Gamma_j^2 - (\omega_j - \omega_g)^2\right)$ and $c_j = 2.f_j.\Gamma_j.(\omega - \omega_g)$, $\boldsymbol{n_\infty}$ is the refractive index and equal to the value of the refractive index when ($\omega \rightarrow \infty$), $f_j$ (eV) and are related to the strength (amplitude) of the peak of the extinction coefficient. $\Gamma_j$(eV) is the broadening term of the peak of absorption. $\omega_j$(eV) is the energy at which the extinction coefficient is maximum. $\omega_g$(eV) is the optical band gap $E_g$

## 3. Results and discussions

### 3.1 Optical properties of films by (UV/vis/NIR) Spectrophotometry analysis

Optical transmission and reflectance spectra for the ZnO and AZO ultrathin layers with the thickness of about 20 nm thick on different glass substrate and PET were measured between (200-2200 nm) are represented in Fig. 2 (a, b) respectively. The optical transmittance spectra clearly exhibit a shift in the band edge due to the aluminum concentration. Also, the transmittance spectra of the glass and PET substrates was measured and illustrated in these figures for comparison. It can be seen; the average transmittance values of ZnO and AZO films in the visible spectrum (400-800 nm) region have been calculated from the spectra of the films. The average visible transmittance of ZnO and AZO films is around 85%-88% on glass and 80%-82% on PET. The difference of the transmittance values can be attributing to the transmittance of the different substrates. The calculated absolute average optical transmittance values $T_{ave}$ of the films can be determined using the expression $T_{absolute} = \frac{T_{film}}{T_{substrate}}$ [22], where $T_{substrate}$ and $T_{film}$ are the transmittance values of the substrate and film respectively. In our case, the absolute $T_{ave}$ of ZnO and AZO ultrathin films are 87%-90% on glass, 82%-85% on PET. This means that the transmittance of AZO films is greater than those of ZnO films and remaining high in the visible region for all the films deposited on glass or on PET substrates. On the other hand, the average reflectance of films in the visible range of the ZnO and AZO films is observed in the Fig. 1 (a and b). We have measured 11% - 9% , 9.9% - 4% on glass and 14%-12%, on PET are shown in the inset of Fig 3 (a) and (b) respectively, the reflectance of AZO films decreased as compared with ZnO pure due to the increasing in the voids of AZO films. The transmittance curves of films are smooth with no oscillation in the visible spectrum meaning the films have low thicknesses. As a result, the optical property of the films on glass is better than films on PET substrates.

As it is known, the absorption coefficient for the films as a function of wavelength $\alpha$ ($\lambda$) calculated according to the optical transmittance values T and thickness of the thin films by the Lambert-Beer expression [23] $\alpha = \frac{-lnT}{t}$, where $T$ and $t$ is the sample's transmittance and thickness respectively. The figure 2 shows the absorption coefficient of films as a function of wavelength of the incident light. The absorption coefficient of the ZnO and AZO ultrathin films is in the spectra of $(10^5$-$10^6)$ $cm^{-1}$ as $hv > E_g$. The absorption coefficients in the UV spectra are larger than those in the visible and near-infrared spectra, which are in the range of $(10^4$-$10^5)$ $cm^{-1}$ [see inset of Fig. 3] and decreases with the Al concentration on both glass and PET substrates as compared with ZnO pure. In addition, the absorption coefficient values on PET in greater than those on glass. The differences of the $\alpha$ values of films in the UV range is due to the effect of the optical energy gap. These results prove that the absorption in the UV range of the AZO films is reduced by Al mergence.

Therefore; knowing the absorption coefficient values $\alpha$ and the relationship with the incident photon energy [$E = hv$, where $h$: planck's constant, $v$: the photon frequency].  The optical band gap of films can be calculated by the Tauc's relationship given by the following equation [24]

$$\alpha = \frac{B(E-E_g)^{\frac{1}{m}}}{E} \qquad (2)$$

where $\alpha$ is absorption coefficient, $B$ is constant, $E_g$ is the band gap, $E$ is the photon energy.  The values $m$ =0.5, 1.5, 2, and 3 depending on the nature of the interband electronic transitions, such as allowed direct (DT), forbidden direct (DFT) and allowed indirect (IDT) and forbidden indirect (IDFT) transitions respectively. By extrapolating the linear part of the plot of $(\alpha E)^{\frac{1}{m}}$ versus the values $E$ (eV), the optical band gap of films is obtained where $(\alpha E = 0)$ in this graph.  The values of $E_g$ due to four optical transitions of ZnO thin film deposited on glass and PET substrates are shown in Fig. 3.  (a and b) on glass substrates and Fig. 4 (c and d) on PET substrates.

The values of optical energy gap of ZnO and AZO films were reported in table 2. As it can be seen, the value of direct band gap is greater than indirect band gap of films deposited on two types of substrates and the optical band gap values of ZnO thin films is smaller than those of AZO films, which is in agreement with the literature.

## 3.2 Optical properties of films by (SE) ellipsometry analysis

The SE measure $\Psi(\lambda)$ and $\Delta(\lambda)$ directly, then the optical properties of the films can be obtained.  The model is fitted and with the help of the DeltaPsi2 software allows to obtain the refractive and extinction indices $n(\lambda)$ and $k(\lambda)$ of the films respectively. The thickness of the glass and PET substrates proposed is nearly infinite as compared to that of ZnO and AZO films. The mean square error $\chi^2$ value is used to appoint the difference between the experimental and theoretical results. This value should be small as low as possible. The $\chi^2$ is defined as follows [25]:

$$\chi^2 = min \sum_1^n \left[ \frac{(\Psi_{th} - \Psi_{exp})_i^2}{\Gamma_{\Psi,i}} + \frac{(\Delta_{th} - \Delta_{exp})_i^2}{\Gamma_{\Delta,i}} \right] \qquad (8)$$

where $\Gamma_i$: is the standard deviation of the points. The smallest value of $\chi^2$ refers to a better fitting results.

For the model structure; the Void layer (represents air) obey Fixed Index dispersion formula was used to describe the optical properties of this layer, this proposed model is given by constant refractive index and extinction coefficient for any wavelength according to the following equations [25] $n(\lambda) = constant = n, k(\lambda) = constant = k$, where $n$ and $k$ is the value of the refractive and extinction index respectively.

For glass layer, a new amorphous dispersion formula is used in order to give a Lorentzian shape to the expressions of the refractive index and extinction coefficient. Equations of $k(\omega)$ and $n(\omega)$ in the section 2.4 in this paper expressed this dispersion formula. The physical parameters of PET layer are listed in table 3.

For PET layer the Tauc-Lorentz dispersion formula is used to describe the physical parameter of this layer. Jellison and Modine developed this model [21] using Tauc joint density of states and Lorentz oscillator. This formula is applied to describe its complex dielectric function $\tilde{\varepsilon}_T = \varepsilon_r + i\varepsilon_i$ where $\varepsilon_r$ and $\varepsilon_i$ is the real and imaginary parts of dielectric constant. The imaginary part of Tauc-Lorentz formula is expressed as follows

$$\varepsilon_r(E) = \varepsilon_r(\infty) + \frac{2}{\pi} P \int_{E_g}^{\infty} \frac{\xi.\varepsilon_i(\xi).d\xi}{\xi^2 - E^2} \text{ and } \varepsilon_i = \begin{cases} \frac{1}{E} . \frac{A_i.E_i.C_i.(E-E_g)^2}{(E^2-E_i^2)^2+C_i^2.E^2}, & for\ E > E_g \\ 0 & , for\ E \leq E_g \end{cases}$$

where $P$: is the Cauchy principal value, the $(i)$ refers to the number of oscillators; in our case $i = 3$. One parameter is linked to the real part of dielectric function, $\varepsilon_r(\infty) = \varepsilon_\infty$ is the high frequency dielectric constant. Four parameters are used to describe imaginary part of the dielectric function. $A_i$ (eV) is related to the strength of the absorption peak. $C_i$(in eV) is the broadening term. $E_g$(eV) is the Tauc's optical band gap energy.

E(eV) is the energy of maximum transition probability. The physical parameters of the PET layer in our case are listed in table 4.

For the ZnO and AZO films layer the spectral dependencies of $\Psi(\lambda)$ and $\Delta(\lambda)$ should modeled until we obtain the best fit with the experimental. The experimental and fitting results of these parameters as a function of wavelength of the ZnO films on glass and PET substrates are depicted in figure 4 (a, b) respectively. From the figure 4 and table 4 it can be seen that we obtain a good agreement of the experimental (line) and theoretical or model fit (dash) data of the films deposited on glass and PET substrates using a new amorphous dispersion formula as a model.

The experimental and fitting results of $\Psi(\lambda)$ and $\Delta(\lambda)$ of the AZO films on glass and PET substrates have been investigated too (curves not shown here). The fitted physical parameters of the new amorphous dispersion formula for the ZnO and AZO samples on glass and PET substrates are listed in table 5.

The experimental and fitting results $n(\lambda)$ and $k(\lambda)$ as a function of wavelength acquired from $\Psi(\lambda)$ and $\Delta(\lambda)$ spectra of the ZnO and AZO films on glass and PET substrates are presented in the figure 5 (a, b) and (c, d) respectively. As it can be seen for all thin films, $n$ and $k$ values on PET substrates are smaller to those on glass substrates. The refractive and extinction coefficients of the AZO films decrease with those of ZnO films due to an increase in the carrier concentration in the case of AZO films. For all of the ZnO films, the $n$ and $k$ values in the visible range is greater than those of AZO films, and a shift in the maximum value of $n$ for AZO films compare with ZnO pure the inset of Fig. 6 (a and c) on glass substrate and Fig. 6 (c and d) on PET substrate. The values of the $n$ and $k$ of films on both substrates sharply changed in the visible region and decreases with the wavelength increases. The decreasing of the refractive index with the Al doping concentration can be chiefly ascribed to an increasing of the carrier concentration in the AZO films. $Al^{+3}$ pentavalent impurity doped into ZnO films can behave as an effective donor due to the substitution of $Al^{+3}$ ions into the $Zn^{+2}$ sites or mergence of $Al^{+3}$ ions in interstitial places, producing free carriers. The carrier concentration in the AZO films is increased with the increasing dopant concentration. Therefore, the refractive index is decreased. Moreover, the extinction coefficient, the real and the imaginary components of dielectric also decrease with the Al doping concentration. The decrease of these parameters with the Al doping concentration may also be ascribed to an increase of the carrier concentration in the AZO films.

The absorption coefficient $\alpha(\lambda)$ of the ZnO and AZO ultrathin films deposited on glass and PET were presented in the figure 6 (a) and (b) respectively. We find that the absorption coefficients in the UV spectra in the range of $10^6$ cm$^{-1}$ are larger than those in the visible and near-infrared spectra in the range of ($10^4$ to $10^5$ cm$^{-1}$) on both glass and PET substrates in the inset of the figure 6 (a) and the figure 6 (b) respectively, and decreased with the Al doping concentration because the glass and PET substrates absorbed the short wavelengths and there are transparent in larger wavelengths. Moreover, those in the ultraviolet region changed little with Al doping concentration (Al incorporation), which may be as a result to the decrease in the optical energy gap. These results reveal that the optical absorption edge in the UV region of the AZO thin films can be decreased by Al doping concentration and the absorption coefficient is a wave number dependent function, which is in a good agreement with those of UV measurements.

In table 6, The direct band gap values of films were obtained by extrapolating the linear region of the curves to the zero absorption at which $(\alpha E)^2 = 0$ and the results are presented in table 6. The plots of $[(\alpha E)^{1/2}$ and $(\alpha E)^{1/3}$ versus E] corresponding to the indirect allowed and forbidden transitions whereas $[(\alpha E)^2$ and $(\alpha E)^{2/3}$ versus E] corresponding to the direct allowed and forbidden transitions. The band gap values of the prepared films corresponding to indirect allowed ($r = 2$) and indirect forbidden transitions ($r = 3$) have been obtained from the linear region of their curves (not shown here). According to Tauc relation it is possible to distinct three regions in the absorption edge spectra. The first region (tail) is the weak absorption, the second region is the exponential edge region $(1 < \alpha < 10^4)$ and the third region $(\alpha > 10^4)$ is the high absorption. It is observed from table 6 that direct optical band gap values are found to be higher than the indirect optical band gap values. Both the direct and indirect optical band gap values are found to decrease with the Al content on different substrates, which indicates that Al content causes structural changes in the films system.

The study of optical edge is useful to understand the optically induced transitions and the optical band gap of the films. The optical absorption spectrum have been used to determine optical band gap values. The optical band gap results of the films on both glass and PET substrates for four electronic transitions have been extracted directly from the DeltaPsi2 software (Tauc Plot) and confirmed mathematically by Tauc's formula equation (2) for each type of optical transitions. The energy gap of the films for different optical transitions modes by SE measurements experimentally($E_{g_{EXP}}$) and theoretically($E_{g_{FIT}}$), on glass and PET substrates (see table 6) are in an excellent agreement with those of UV values.

For comparison according to table 1 and table 6 we obtain an excellent agreement between the experimental and fitting results of the optical band gap values for ZnO and AZO films deposited on both glass and PET substrates measured by SE for the four types of electronic transitions and with the UV method.

## 4. Conclusion

ZnO and AZO ultrathin films deposited on two types of substrates at 300 K by DC sputtering technique. XRD patterns of the films coated on both glass and PET substrates show an amorphous structure phase. The optical properties of the deposited films on glass and PET substrates show a transmittance of 80% to more than 90% in the visible spectrum in both methods UV and a new amorphous model in the wavelength range of 200-2200 nm. The absorption coefficient and optical band gap values of films on both glass and PET substrates by UV method showed an excellent agreement with those of SE measurements.

## Acknowledgement

Authors are grateful to Nicolas Mercier, Magali Allain for providing the necessary facilities for XRD studies, Also, to Jean-Paul Gaston and Celine Eypert from Jobin Yvon Horiba Company for the spectroscopic ellipsometry measurements and to Cecile Mézière, Valerie BONNIN for help with the chemicals and corresponding equipment.

**Figures Captions**

Fig. 1. (a): Optical transmittance and reflectance curves of ZnO and AZO ultrathin films on glass and (b): on PET substrates.

Fig. 2 : Absorption coefficient curves of ZnO and AZO films deposited on glass and PET substrates.

Fig. 3. (a and b): the optical band gap of ZnO films for different optical transitions modes deposited on glass and (c and d) on PET substrates.

Fig. 4. (a and b): Experimental and fitted spectra of $\Psi(\lambda)$ and $\Delta(\lambda)$ of ZnO on glass and on PET substrates respectively.

Fig. 5: (a and b): Experimental and fitting results of the n($\lambda$) and k($\lambda$) curves of ZnO and AZO films on glass and (c and d) on PET substrates.

Fig. 6. (a): Experimental and fitting results of α($\lambda$) curves of the ZnO and AZO films on glass substrate and (b): on PET substrate.

# List of tables

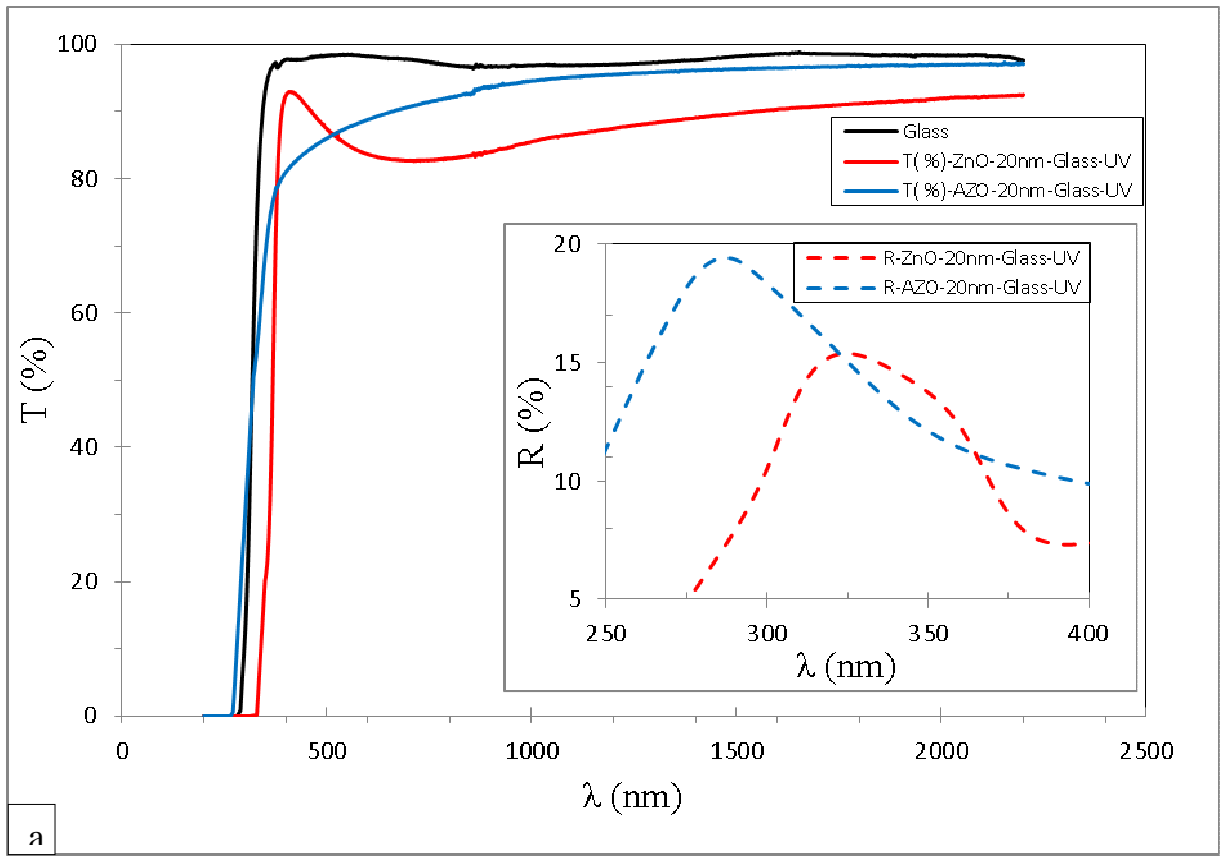

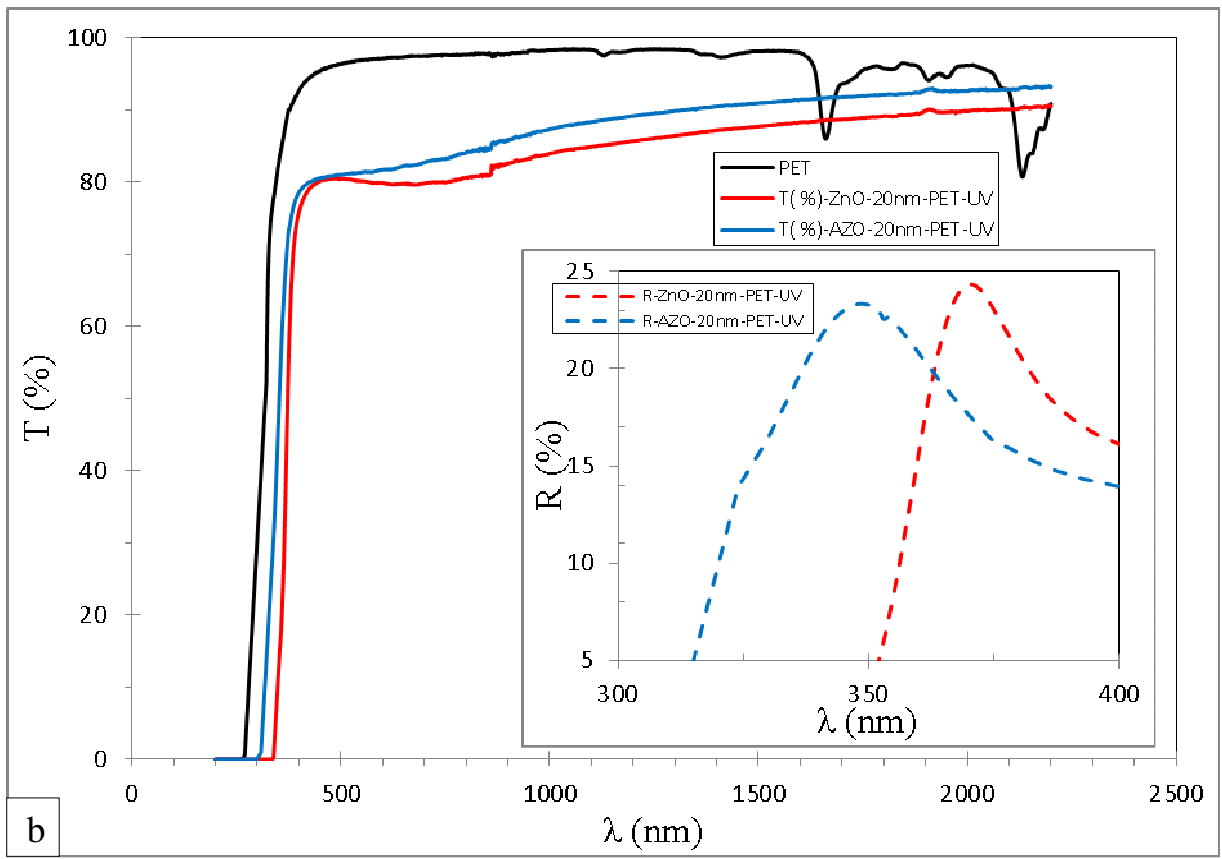

Fig. 1

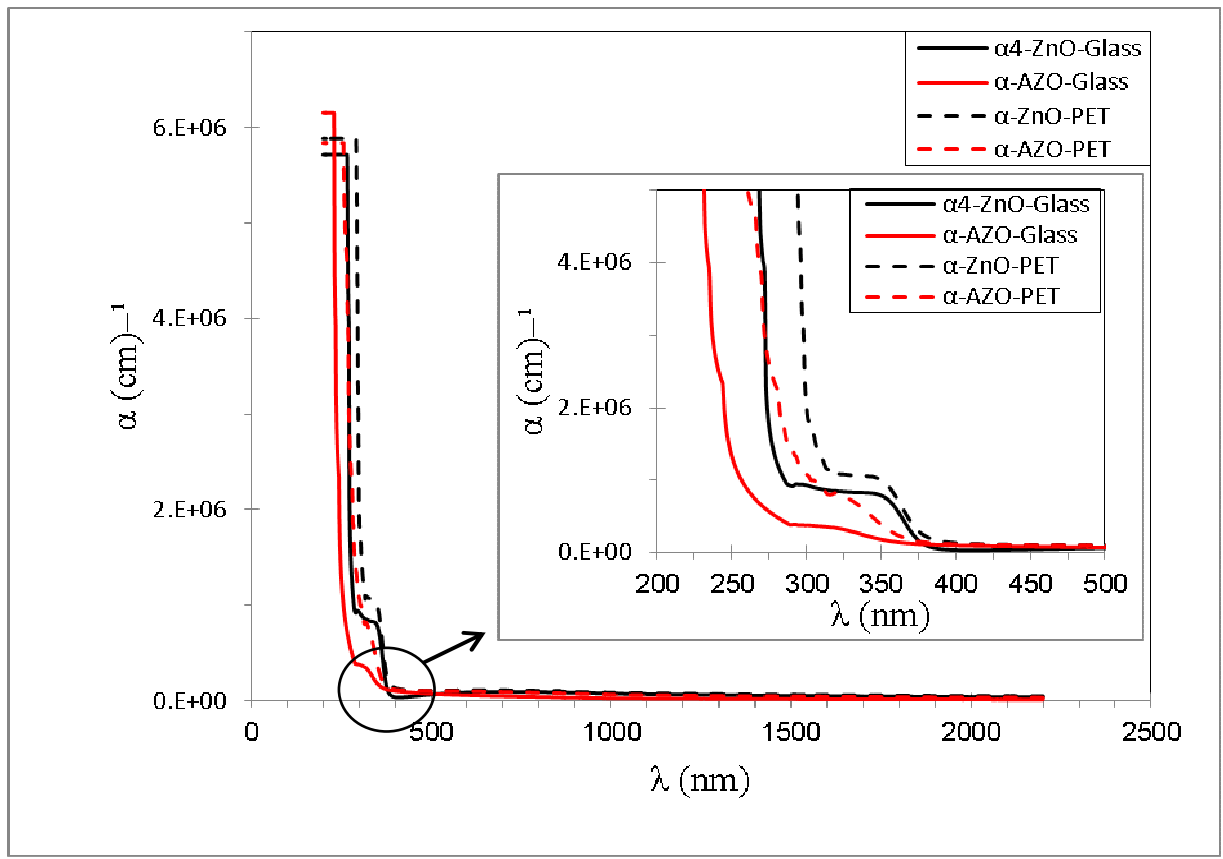

Fig. 2

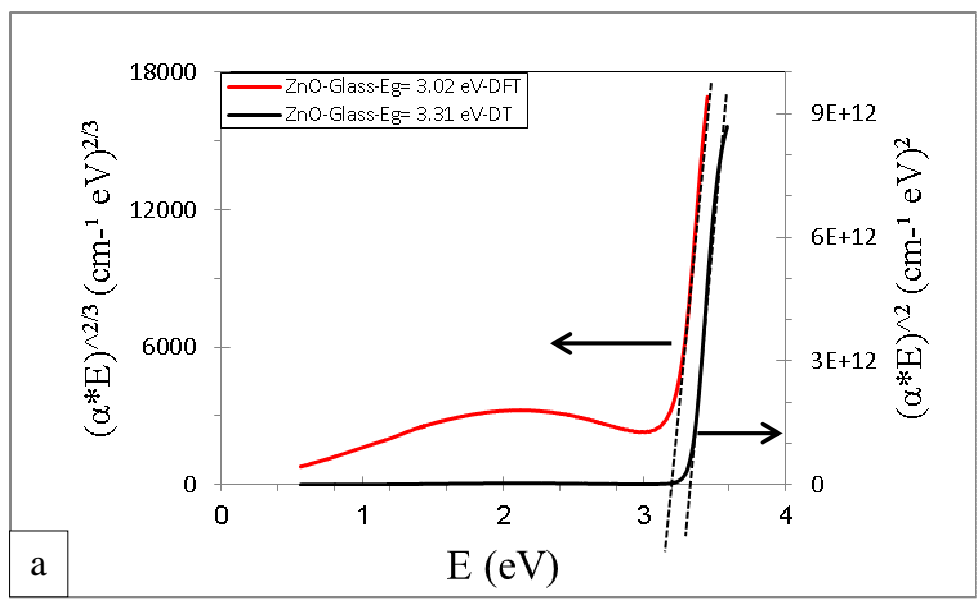

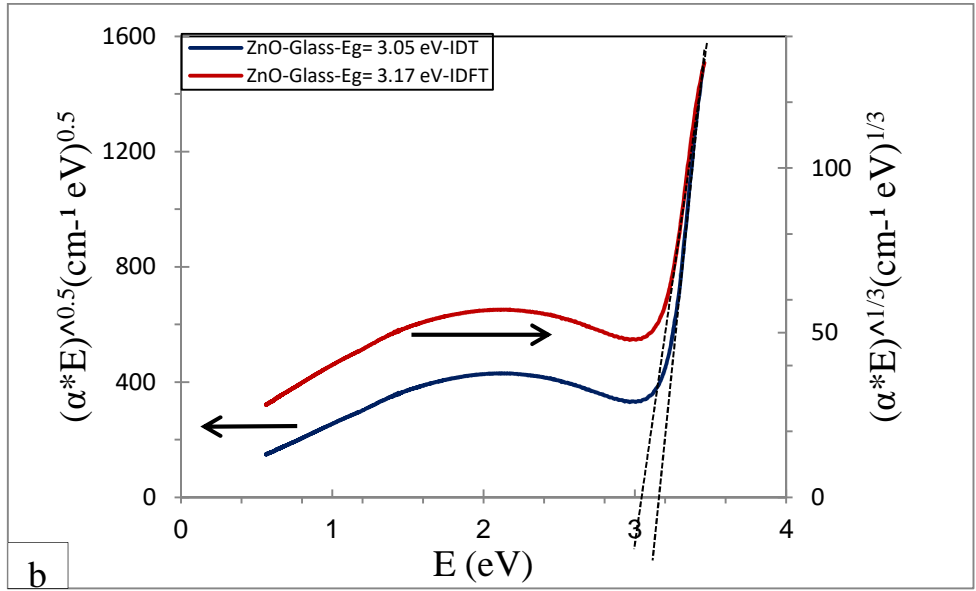

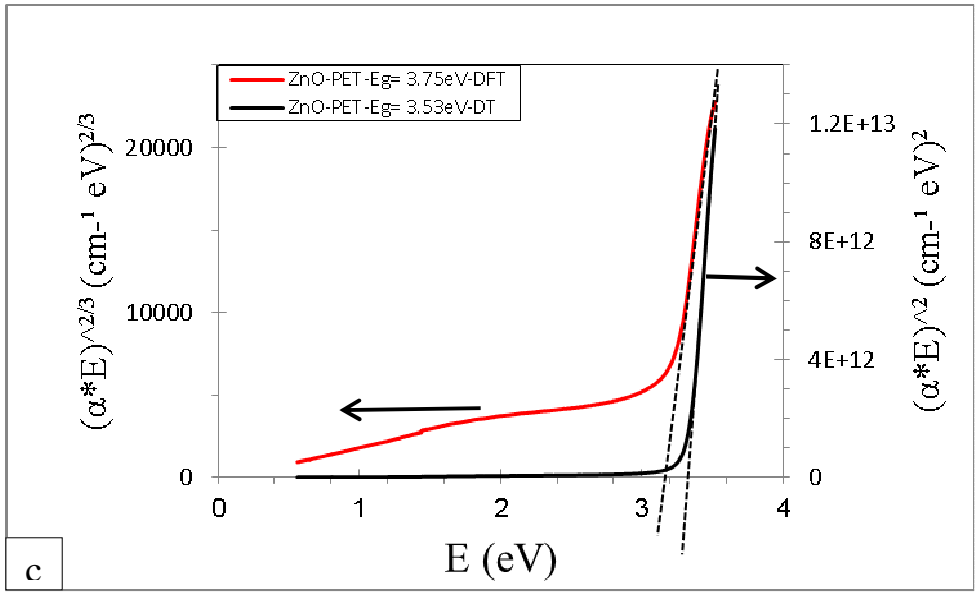

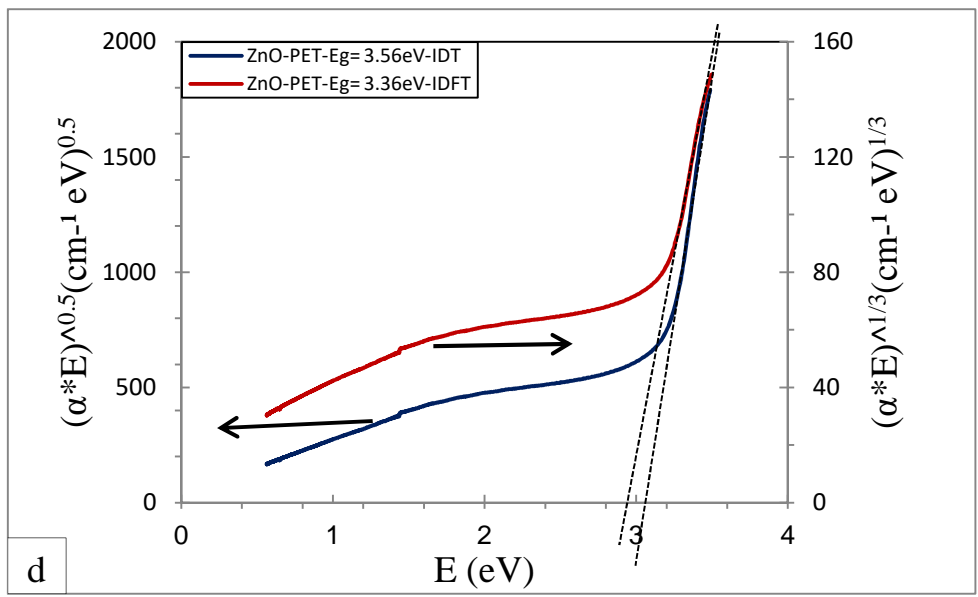

Fig. 3

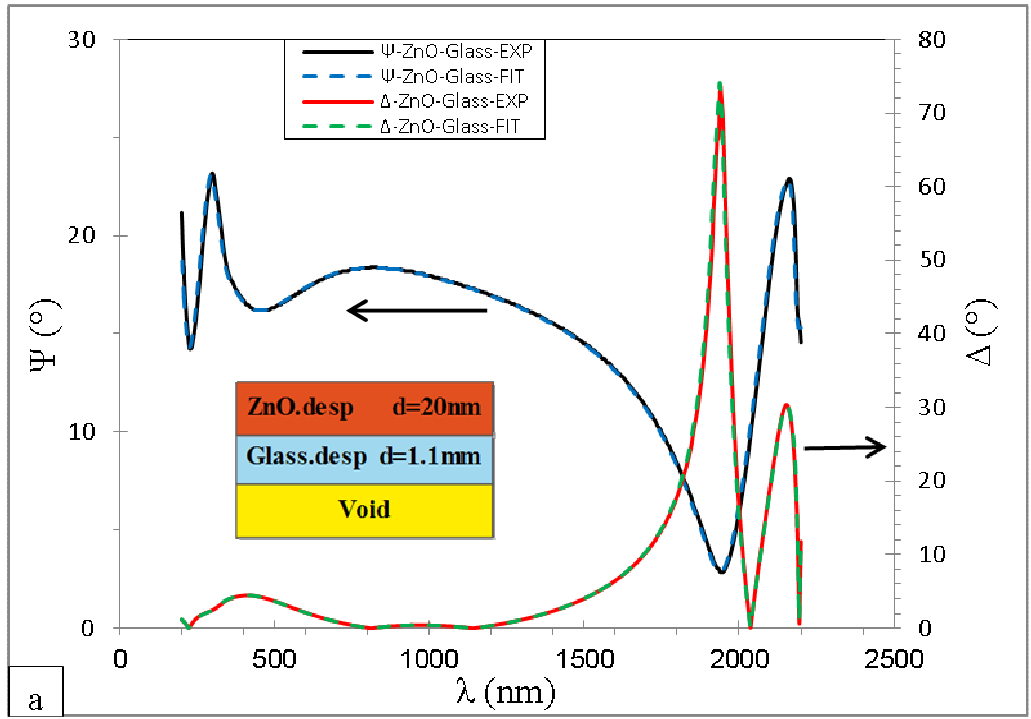

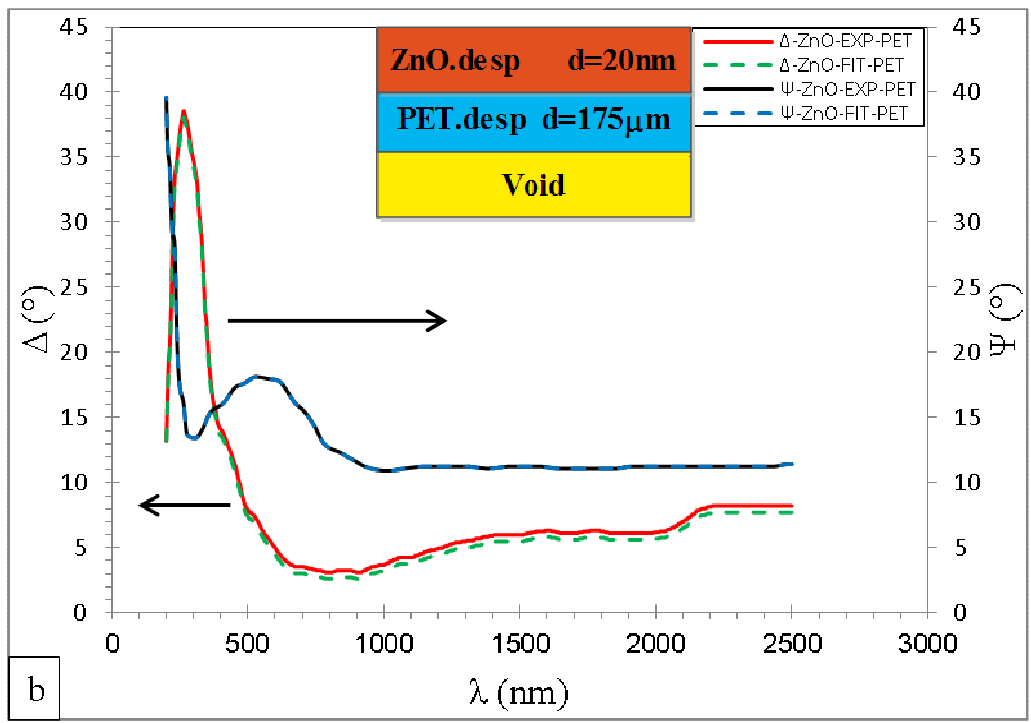

Fig. 4

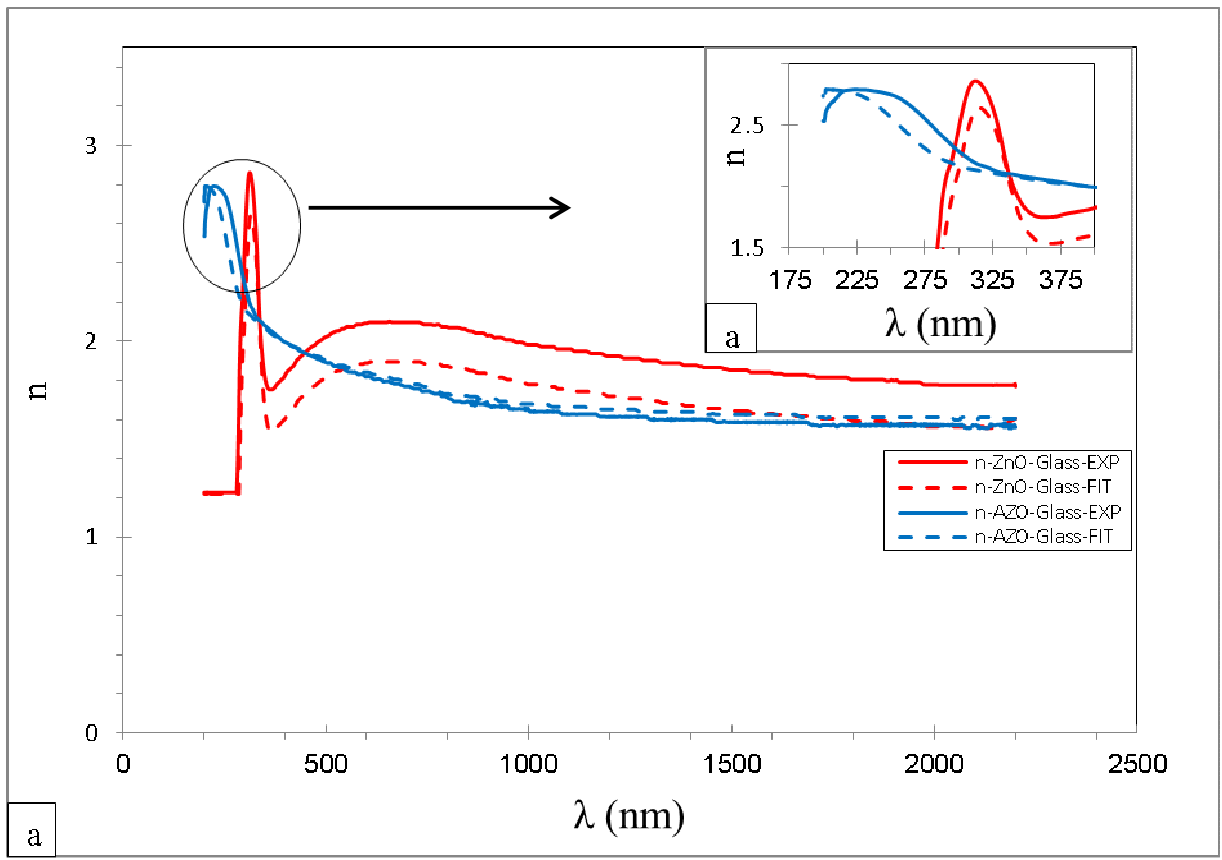

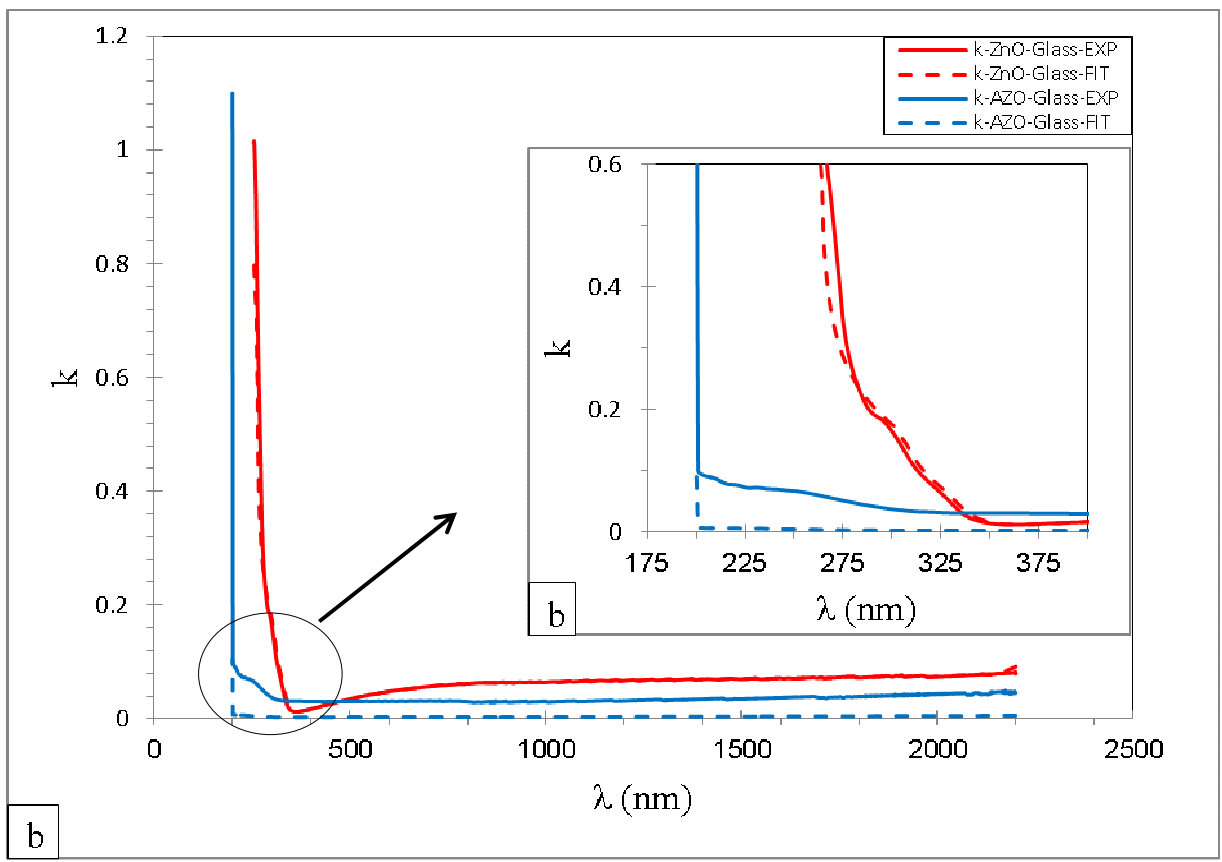

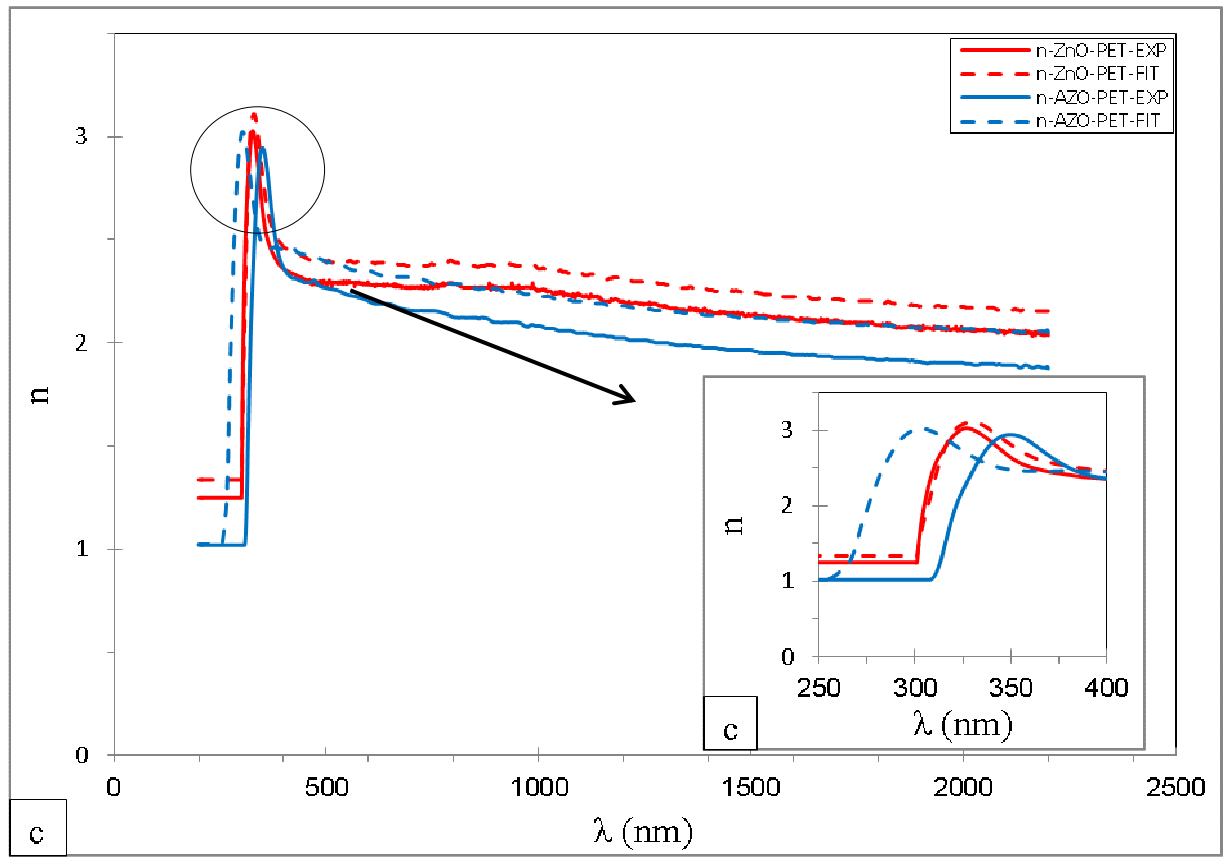

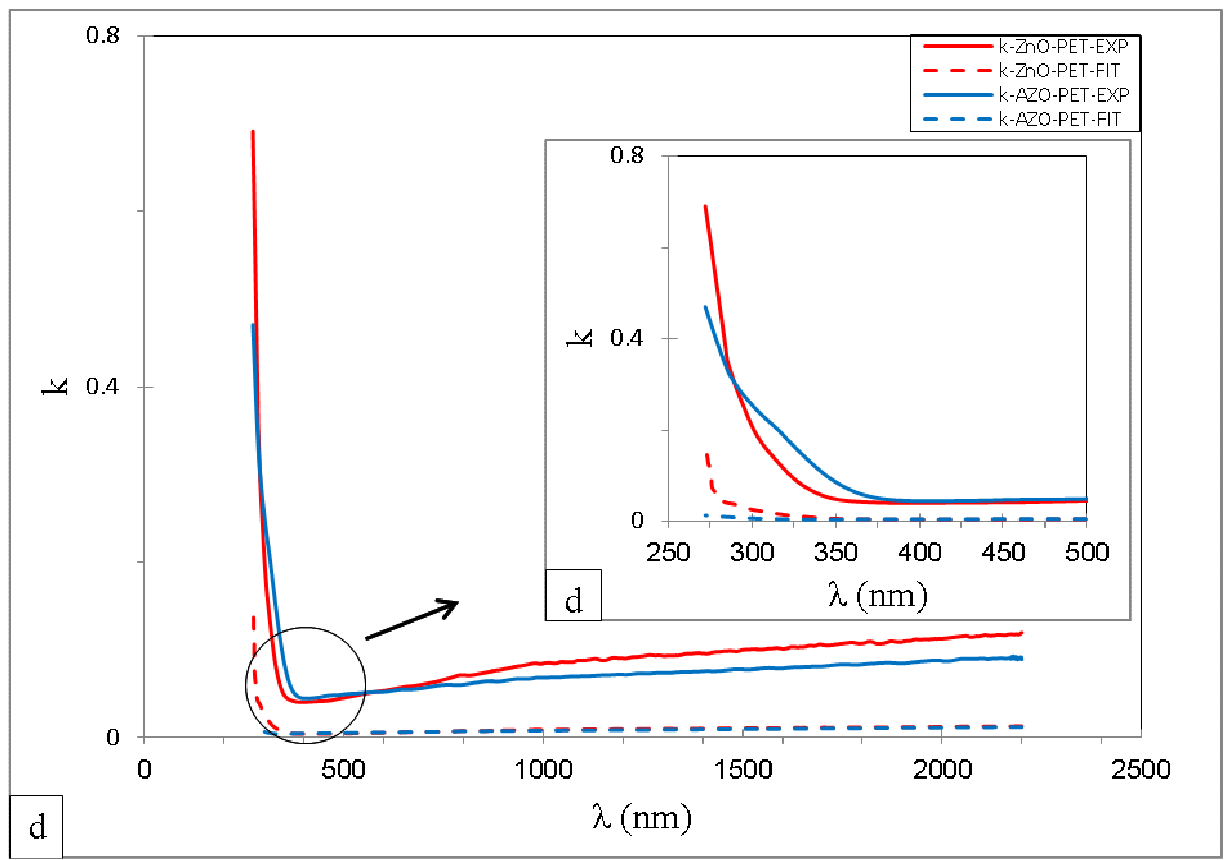

Fig. 5

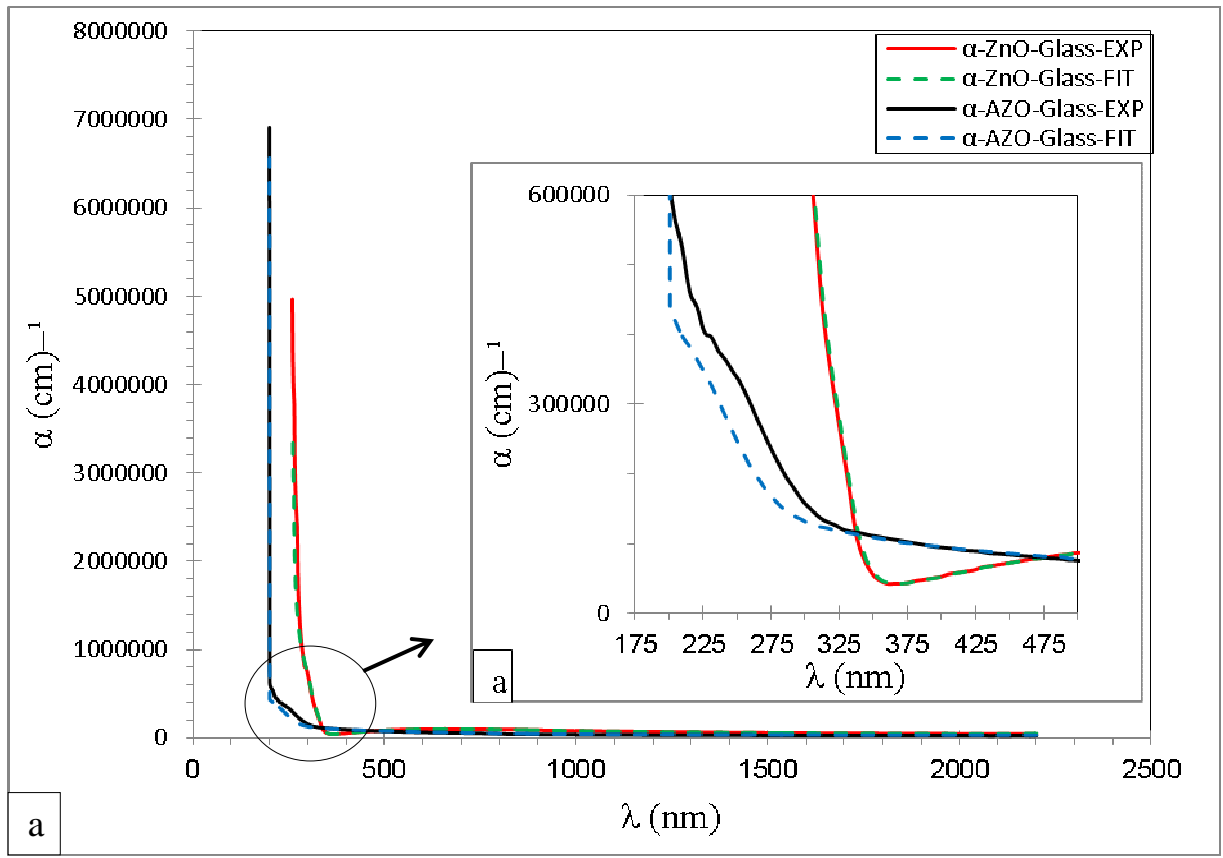

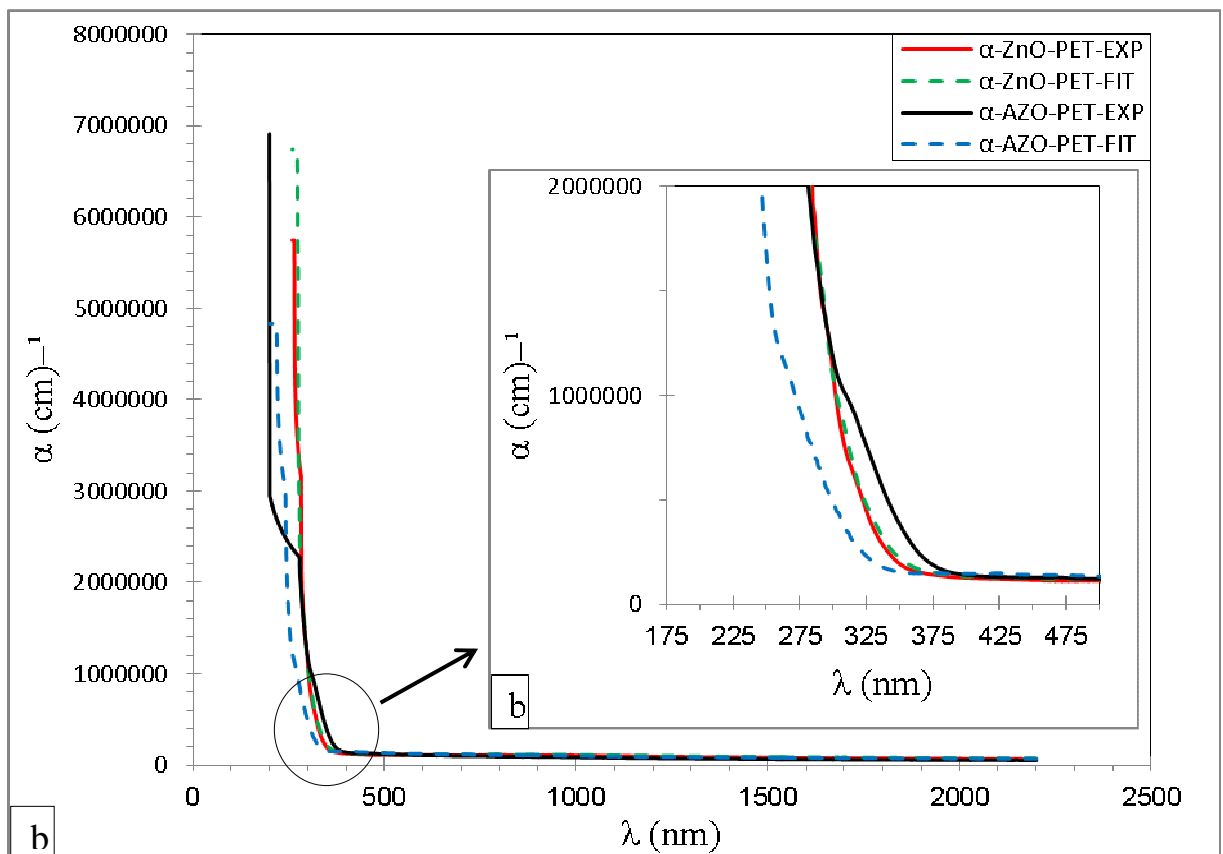

Fig. 6

| Samples | Profilometer | | Ellipsometry | |
|---|---|---|---|---|
| | Thickness (nm) | | | |
| | Glass | PET | Glass | PET |
| ZnO | 20.9 | 20.8 | 20.8 | 20.4 |
| AZO | 20.7 | 20.8 | 20.4 | 20.9 |

Table 1

| | Thickness (nm) | Spectrophotometry measurements | | | | | | | |
|---|---|---|---|---|---|---|---|---|---|
| Substrate | | Glass | | | | PET | | | |
| Optical transition | | DT | DFT | IDT | IDFT | DT | DFT | IDT | IDFT |
| ZnO | 20 | 3.31 | 3.02 | 3.05 | 3.17 | 3.5 | 3 | 3.3 | 2.95 |
| AZO | 20 | 3.53 | 3.15 | 3.5 | 3.36 | 3.6 | 3.5 | 3.58 | 3.4 |

Table 2

| Model parameters | Glass substrate | | | | | | |
|---|---|---|---|---|---|---|---|
| | $n_\infty$ | $\omega_g$ | $f_j$ | $\omega_j$ | $\Gamma_j$ | $AOI$ | $d$ (nm) |
| | 0.626 | 3.28 | 0.055 | 41.765 | 2.309 | 70 | 1100000 |

Table 3

| Model parameters | PET substrate | | | | | | | | | | | |
|---|---|---|---|---|---|---|---|---|---|---|---|---|
| | $\varepsilon_\infty$ | $E_g$ | $A_1$ | $E_1$ | $C_1$ | $A_2$ | $E_2$ | $C_2$ | $A_3$ | $E_3$ | $C_3$ | AOI | d (nm) |
| | 2.5 | 3.7 | 6.4 | 4.2 | 0.2 | 22.2 | 5 | 0.6 | 8.7 | 6.2 | 0.2 | 70 | 175*10³ |

Table 4

| Model parameters | Glass | | PET | |
|---|---|---|---|---|
| | ZnO | AZO | ZnO | AZO |
| $\chi^2$ (MSE) | 0.0005 | 0.167 | 0.219 | 0.239 |
| $n_\infty$ | 1.380±0.001 | 4.925±0.017 | 3.026±0.028 | 10.408±0.232 |
| $\omega_g$ | 1.146±0.001 | 9.251±0.022 | 7.876±0.106 | 28.159±0.204 |
| $f_j$ | 0.001±0.005 | 1.407±0.002 | 0.420±0.006 | 54.181±6.729 |
| $\omega_j$ | 0.493±0.001 | 0.742±0.007 | 2.259±0.027 | 31.009±0.173 |
| $\Gamma_j$ | 0.012±0.001 | 4.998±0.009 | 3.157±0.052 | 9.522±0.889 |
| $AOI$ | 72.479±0.000 | 70.994±0.023 | 72.780±0.003 | 72.604±0.003 |
| $d$ (nm) | 20.876±0.007 | 20.414±0.017 | 20.403±0.033 | 20.925±0.050 |

Table 5

| Optical band gap calculations by spectroscopic ellipsometry | | | | | | | | | |
|---|---|---|---|---|---|---|---|---|---|
| Materials | Thickness (nm) | Glass substrate | | | | | | | |
| | | DT | | DFT | | IDT | | IDFT | |
| | | $E_{g_{EXP}}$ | $E_{g_{FIT}}$ | $E_{g_{EXP}}$ | $E_{g_{FIT}}$ | $E_{g_{EXP}}$ | $E_{g_{FIT}}$ | $E_{g_{EXP}}$ | $E_{g_{FIT}}$ |
| ZnO | 20.876 | 3.31 | 3.31 | 3.02 | 3.02 | 3.05 | 3.05 | 3.17 | 3.17 |
| AZO | 20.414 | 3.53 | 3.53 | 3.15 | 3.15 | 3.5 | 3.5 | 3.36 | 3.36 |
| Materials | Thickness (nm) | PET substrate | | | | | | | |
| | | $E_{g_{EXP}}$ | $E_{g_{FIT}}$ | $E_{g_{EXP}}$ | $E_{g_{FIT}}$ | $E_{g_{EXP}}$ | $E_{g_{FIT}}$ | $E_{g_{EXP}}$ | $E_{g_{FIT}}$ |
| ZnO | 20.403 | 3.5 | 3.5 | 3 | 3 | 3.3 | 3.3 | 2.95 | 2.95 |
| AZO | 20.925 | 3.6 | 3.6 | 3.5 | 3.5 | 3.58 | 3.58 | 3.4 | 3.4 |

Table 6